\documentclass[fleqn,twoside]{article}
\usepackage{espcrc2}

\usepackage{graphicx,epsfig}
\usepackage[figuresright]{rotating}

\newcommand{\AmS}{{\protect\the\textfont2
  A\kern-.1667em\lower.5ex\hbox{M}\kern-.125emS}}

\title{Initial State Radiation: A success story}

\author{W. Kluge\address{Institut f\"{u}r Experimentelle
Kernphysik, Universit\"{a}t Karlsruhe, \\Postfach 3640, D 76021
Germany}\thanks{My thanks for the fruitful cooperation in more
than 15 years go to the \emph{KLOE} collaboration, in particular
to A. Denig, S. M\"{u}ller and G. Venanzoni, to F. Jegerlehner, J.
H. K\"{u}hn and H. Czy\.{z} for many enlightening discussions, to
G. Pancheri for their guidance of the European networks EURODAFNE
and EURIDICE and to the Laboratori Nazionali di Frascati for the
generous hospitality. The experiment \emph{KLOE} was supported in
part by EURIDICE contract HPRN-CT2002-00311; by the German Federal
Ministry of Education and Research (BMBF) contract 06-KA-957; by
Deutsche Forschungsgemeinschaft, contracts KL 1820/1-1 and KL
1820/1-2; and by the EU Integrated Infrastructure Initiative
HadronPhysics Project (TARI) under contract number
RII3-CT-2004-506078.}}

\begin{document}

\begin{abstract}
The investigation of events with \emph{Initial State
Radiation(ISR)} and subsequent \emph{Radiative Return} has become
an impressively successful and guiding tool in low and
intermediate energy hadron physics with electron positron
colliders: it allows to measure hadronic cross sections and the
ratio \emph{R} from threshold up to the maximum energy of the
colliders running at fixed energy, to clarify reaction mechanisms
and reveal substructures (intermediate states and their decay
mechanisms) and to search for new highly excited mesonic states
with $J^{PC} = 1^{--}$. While being discussed since the
sixties-seventies \emph{ISR} became a powerful tool for
experimentalists only with the development of \emph{EVA-PHOKHARA}
\cite{Binner,Czyz}, a Monte Carlo generator developed over almost
10 years, while increasing its complexity, which is user friendly,
flexible and easy to implement into the software of existing
detectors. \vspace{1pc}
\end{abstract}

\maketitle

\section{INITIAL STATE RADIATION:
\newline
The idea} The idea to use \emph{Initial State Radiation} to
measure hadronic cross sections from the threshold up to the
maximum energy of colliders with fixed energies $\sqrt{s}$, to
reveal reaction mechanisms and to search for new mesonic states
consists in exploiting the process $e^+e^-\rightarrow hadrons + n
\gamma$ to reduce the centre of mass energies of the colliding
electrons and positrons and consequently the energy squared
$M^2_{hadr}= s - 2 \sqrt{s} \: \: E_{\gamma}$ of the final state
by emitting 1 or more photons. The method is particularly well
suited for the modern meson factories like $DA\Phi NE$ running at
the $\phi$-resonance, \emph{PEP-II}, \emph{KEKB} at the $
\Upsilon(4S)$-resonance with their high luminosities which
compensate for the $\alpha / \pi$ suppression of the emission of a
photon. $DA\Phi NE$, \emph{PEP-II} and \emph{KEKB} cover the
energy regions up to 1.02 GeV and up to 10.6 GeV, respectively
(restricted for the latter actually up to 4...5 GeV if hard
photons are detected). A big advantage of the \emph{ISR} method
are the low point-to-point systematic errors of the hadronic
energy spectra because the luminosity, the energy of the electrons
and positrons and many of the various contributions to the
detection efficiencies are determined once for the whole spectrum.
As a consequence the overall normalization error is the same for
all energies of the hadronic system. The term \emph{Radiative
return} alternately used for \emph{ISR} refers to the appearance
of pronounced resonances (e.g. $\rho, \omega, \phi, J/\psi, Z$)
with energies below the collider energy (see Fig.
\ref{fig:Rsigma.eps}, taken from Ref. \cite{PDG}). Reviews and
updated results can be found in the Proceedings of the
International Workshops in Pisa (2003) \cite{Pisa}, Novosibirsk
(2006) \cite{Budker} and the present ones in Frascati (2008).

\section{INITIAL STATE RADIATION:
\newline The history}
\subsection{The pre-\emph{EVA-PHOKHARA} era}
\begin{figure}[htb]
\includegraphics[width=70mm]{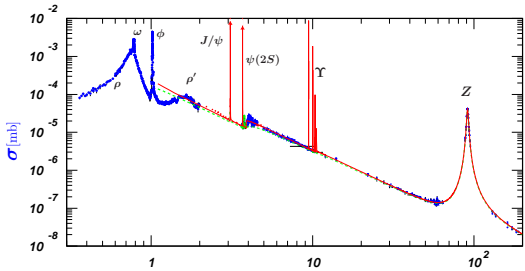}
\vspace{0.5cm}
\vglue -1.0cm
\caption{Initial State Radiation and the cross
section $\sigma$ as function of centre of mass energy $\sqrt{s}$,
courtesy PDG Ref. \cite{PDG}} \label{fig:Rsigma.eps}
\end{figure}

Calculations of \emph{ISR} date back to the sixties-seventies of
the $20^{th}$ century. For example photon emission for muon pair
production in electron-positron collisions has been calculated in
Ref. \cite{Baier1}, for the $2 \pi $-final state in Ref.
\cite{Baier2}, resonances ($\rho,\omega,\phi$) have been
implemented in Ref. \cite{Pancheri}, the excitation of
$\psi(3100)$ and $\psi^{\prime}(3700)$ in Ref. \cite{Greco}, and
the possibility to determine the pion form factor was discussed in
Ref. \cite{Zerwas}. The application of \emph{ISR} to the new high
luminosity meson factories, originally to determine the hadronic
contribution to vacuum polarization, more specifically the pion
form factor, has materialized in the late nineties. It has started
with calculations of \emph{ISR} for the colliders $DA\Phi NE$,
\emph{PEP-II}, \emph{KEKB} \cite{Spagnolo,Khoze,Benayoun}. See
also Refs. \cite{Arbuzov1,Arbuzov2} for calculations of radiative
corrections for pion and kaon production below energies of 2 GeV.
An impressive example of \emph{ISR} is the \emph{Radiative Return}
to the region of the \emph{Z}-resonance at \emph{LEP 2} with
collider energies around 200 GeV \cite{OPAL}, (see Fig.
\ref{fig:Rsigma.eps}).

\vglue -0.5 cm

\vspace{-0.8cm}

\subsection{\emph{EVA}, \emph{PHOKHARA}: determination of hadronic cross sections,
the muon magnetic anomaly and the running fine structure constant}

\emph{ISR} became a powerful tool for the analysis of experiments
at low and intermediate energies only with the development of
\emph{EVA-PHOKHARA}, a Monte Carlo generator which is user
friendly, flexible and easy to implement into the software of the
existing detectors \cite{Binner,Czyz}. The generator was applied
to an experiment to determine the cross section $e^+e^-\rightarrow
\pi^{+} \pi^{-} \gamma$ from the reaction threshold up to the
maximum energy for the first time with the detector \emph{KLOE} at
$DA\Phi NE$ \cite{Cataldi,KLOE talks,KLOE paper}. The original
motivation was and is still the determination of the hadronic
contribution to the vacuum polarization from hadronic cross
sections.

\subsubsection{The hadronic contribution to the anomalous magnetic
moment of the muon $a_{\mu}^{had}$}

The determination of the hadronic contribution to the vacuum
polarization, which arises from the coupling of virtual photons to
quark-antiquark-pairs  $ \gamma ^ \star \rightarrow q \bar {q}
\rightarrow \gamma ^\star $, is possible by measuring the cross
section of electron positron annihilation into hadrons with a real
photon emitted in the initial state $e^+e^- \rightarrow \gamma
\gamma^* \rightarrow hadrons \: \gamma$. It is of great importance
for the interpretation of the precision measurement of the
anomalous magnetic moment of the muon $a_\mu$ in $Brookhaven \: (E
821)$ \cite{Brookhaven} and for the determination of the value of
the running fine structure constant at the $Z^o$ resonance $\alpha
(m_{Z}^{2})$, contributing to precision tests of the standard
model of particle physics, see for details \emph{Jegerlehner}
\cite{Jegerlehner}, also this conference \cite{Teubner}.   The
\emph{Novosibirsk} groups $CMD-2$ and $SND$
\cite{cmd2,snd,Novosibirsk2008} pioneered the measurements of
hadronic cross sections by changing the collider energy
(\emph{energy scan}). The \emph{ISR} method represents an
alternate, independent and complementary way to determine those
cross sections with different systematic errors.

\subsubsection{\emph{KLOE}: Determination of $e^{+} e^{-}\rightarrow \pi^{+} \pi^{-} \gamma$
by emission of photons in the initial state, radiative return to
the resonances $\rho$ and $\omega$}

The hadronic vacuum polarization below 2.5 GeV can be determined
only experimentally because calculations within perturbative
\emph{QCD} are unrealistic, calculations on the lattice are not
yet available with necessary accuracy, and calculations in the
framework of chiral perturbation theory are restricted to values
close to the $2 \pi$-threshold. With the maximum energy of 1.02
GeV of $DA\Phi NE$ almost 70 \% of the hadronic correction can be
determined due to the dominance of the $\rho-$resonance. This
correction was previously based on the measurement of the
$e^+e^-\rightarrow \pi^{+} \pi^{-}$ reaction of the
\emph{Novosibirsk} groups \cite{cmd2,snd} by an energy scan.

The \emph{KLOE} collaboration has carried out two analyses: In the
so called small photon angle analysis the cross section
$e^+e^-\rightarrow \pi^+ \pi^- \gamma$ has been measured between
0.6 and 1.0 GeV with a statistical error of less than 0.2 \% and a
systematical error of 1.2 \%. Two pions have been detected in the
drift chamber in the angular interval $50^{\circ} < \theta_{\pi} <
130^{\circ}$ with excellent momentum resolution, disregarding the
photon. However, an additional constraint for the missing photon
angle $\theta_{\gamma} = \theta_{(\vert \vec p (\pi^{-})+\vec p
(\pi^{+}) \vert)} < 15^{\circ}, \: \theta_{\gamma} > 165^{\circ}$
has been required. As a consequence of this restricted phase space
only the $M^2_{\pi\pi}$ region between 0.3 and 1.0 $GeV^2$ is
populated \cite{KLOE talks,KLOE paper,KLOE}. The advantage of the
small angle analysis is negligible background from resonant
($\phi$ -)decays ($e^+e^-\rightarrow \phi \rightarrow \pi^{+}
\pi^{-} \pi^{o}$, $e^+e^- \rightarrow \phi \rightarrow f_{o}
\gamma \rightarrow \pi^{+} \pi^{-} \gamma)$ and from final state
radiation.

In the large photon angle analysis the cross section
$e^+e^-\rightarrow \pi^+ \pi^- \gamma$ can be measured down to the
$2 \pi$- threshold. In addition to the two pions also photons are
detected in the angular interval $50^{\circ} < \theta_{\gamma} <
130^{\circ}$ with an energy of $E_{\gamma} > 50 \: MeV$. This
analysis is hindered by large resonant background and final state
radiation which can be overcome by reducing the energy of $DA\Phi
NE$ below the $\phi$-resonance \cite{Seoul}.

\begin{figure}[htb]
\begin{center}
\includegraphics[width=70mm]{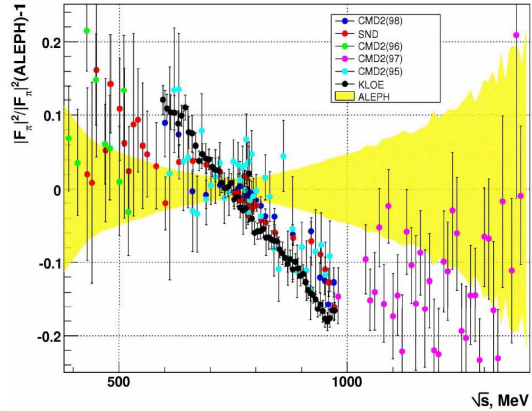}
\vglue -0.5cm \caption{Comparison of data from \emph{KLOE}
\cite{KLOE paper}, \emph{CMD2, SND}
\cite{cmd2,snd,Novosibirsk2008}}
\label{fig:KLOECMD2}
\end{center}
\end{figure}

The results of both analyses agree very well. The hadronic
correction to the vacuum polarization for $a_{\mu}$ obtained in
the small angle analysis using the data of 2001 ($140\: pb^{-1}$
in the interval $0.35 < M_{\pi\pi}^{2} < 0.95 \:GeV^{2})$ is
$a_{\mu}^{\pi\pi}= (388.7 \pm 0.8_{stat} \pm 4.9_{syst})\cdot
10^{-10}$ \cite{KLOE paper}. The total systematic error of
$a_{\mu}^{\pi\pi}$ of $1.3\:\%$ includes an experimental
systematic error of $0.9 \:\%$ and a theoretical systematical
error of $0.9 \:\%$ taken in quadrature. An update (mainly due to
a revised \emph{Bhabha} cross section to determine the luminosity
\cite{Calame}) resulted in $a_{\mu}^{\pi\pi} = (384.4 \pm
0.8_{stat} \pm 4.9_{syst})\cdot 10^{-10}$. Using the data of 2002
($240 \: pb^{-1}$) the small angle analysis gave $a_{\mu}^{\pi\pi}
= (386.3 \pm 0.86_{stat} \pm 3.9_{syst})\cdot 10^{-10}$ with a
noticeably reduced statistical and systematic error. In the large
angle analysis a value of $a_{\mu}^{\pi\pi}= (252.5 \pm 0.6_{stat}
\pm 2.0_{syst} \pm 3.1_{syst,f_{o}})\cdot 10^{-10}$ is obtained
for a restricted region $0.50 < M_{\pi\pi}^{2} < 0.85 \:GeV^{2}$
covering mainly the $\rho$-resonance to be compared with the
result of the small angle analysis of the 2002 data in the same
energy interval of $a_{\mu}^{\pi\pi}= (255.4 \pm 0.4_{stat} \pm
2.5_{syst}) \cdot 10^{-10}$ \cite{Seoul}. The main problem of the
large angle analysis is the large systematic (model dependent)
error from the contribution of the $f_{o}$ from the radiative
decay of the $\phi$ according to $e^+e^- \rightarrow \phi
\rightarrow f_{o} \gamma \rightarrow \pi^{+} \pi^{-} \gamma$ at
low values of $M^2_{\pi\pi}$. The \emph{KLOE ISR} data and the
(energy scan) results from \emph{Novosibirsk}
\cite{cmd2,snd,Novosibirsk2008} agree within the error bars (Fig.
\ref{fig:KLOECMD2}).

\subsubsection{\emph{BaBar}, \emph{Belle}: Determination of $e^+e^-\rightarrow
hadrons \: \gamma$ by emission of photons in the initial state,
radiative return to $\rho, \omega, \phi,..., J/\psi,...$}

Soon after the application of \emph{EVA-PHOKHARA} to \emph{KLOE}
\cite{Cataldi} the \emph{BaBar} collaboration also started the
investigation of hadron final states with $ISR$ \cite{Solodov}. In
recent years a plethora of final states has been studied, starting
with the reaction $e^+e^- \rightarrow J/\psi \: \gamma \rightarrow
\mu^{+} \mu^{-}\: \gamma$ \cite{muplusmuminus}. While detecting a
hard photon the upper energy for the hadron cross sections is
limited to roughly 4.5 GeV. Final states with 3, 4, 5, 6 charged
and neutral pions , 2 pions and 2 kaons, 4 kaons, 4 pions and 2
kaons, with a $\phi$ and a $f_{o}(980)$, $J/\psi$ and 2 pions or 2
kaons, pions and $\eta$, kaons and $\eta$, but also baryonic final
states with protons and antiprotons, $\Lambda^{o}$ and $\bar
\Lambda^{o}$, $\Lambda^{o}$ and $\bar {\Sigma^{o}}$, $\Sigma^{o}$
and $\bar {\Sigma^{o}}$, $D$ and $\bar{D}$-mesons, etc. have been
investigated \cite{BaBar,BaBar2008}. In preparation are final
states with 2 pions, 2 kaons. Particularly important final states
are those with 4 pions which contribute significantly to the muon
anomalous magnetic moment and which were poorly known before the
\emph{ISR} measurements \cite{BaBar2008}. The \emph{BaBar} results
for the $4 \pi$- final states are not only consistent with
previous ones but rather represent the best measurements for
$E_{cm} < 0.75\: GeV$, are competitive for $0.75 \div 1.4 \: GeV$,
are the best ones for $1.4 \div 2.0 \: GeV$ and the only
measurement for $E_{cm}> 2.0\: GeV$. Detailed analyses allow the
identification of intermediate states and consequently the study
of the reaction mechanisms. For instance, in the case of the final
state with 2 charged and 2 neutral pions $(e^+ e^- \rightarrow
\pi^{+} \pi^{-} \pi^{o} \pi^{o} \gamma)$ the intermediate states
$\omega \pi^{o}$ and $a_{1}(1260) \pi $ dominate, while also
$\rho^{+} \rho^{-}$ and $\rho^{o} f_{o}(980)$ have been seen. The
final state with 5 pions proceeds to about $20 \%$ via $\pi^{+}
\pi^{-} \eta$ or $ \rho \eta$ and to about $40 \%$ via $\pi^{+}
\pi^{-} \omega$, the rest being $\rho^{o} \rho^{\pm} \pi^{\mp}$.
For more details see \cite{BaBar2008}.

More recently also \emph{Belle} joined the \emph{ISR} programme
with emphasis on final states containing mesons with hidden and
open charm: $J/\psi$ and $\psi(2S)$, $D$ and $\bar{D}$
\cite{Belle1,Belle2008}.


\subsection{\emph{EVA-PHOKHARA} and hadron spectroscopy}
\subsubsection{\emph{KLOE}: light scalar mesons \newline $\sigma(500)$, $f_{o}(980), a_{o}(980)$}

A long standing problem is the nature of the low lying scalar
mesons $\sigma(500), f_{o}(980), a_{o}(980)$ and their decay into
pseudoscalars.  \emph{KLOE} has studied the production of
$f_{o}(980)$ in the reaction $e^+ e^- \rightarrow f_{o} \gamma
\rightarrow \pi^{+} \pi^{-} \gamma$ \cite{KLOEpiplus}. To separate
the $f_{o}(980)$ from the large irreducible background of initial
and final state radiation the latter has been calculated by
\emph{PHOKHARA}. A large coupling of the $\phi$ to $f_{o}(980)$
has been observed and no conflict found with the $q q \bar{q}
\bar{q}$ hypothesis, see also \cite{Achasov}.

\subsubsection{\emph{BaBar}, \emph{Belle}: New meson states}
Many new highly excited states (preliminarily denoted as \emph{X,
Y, Z}) have been found in recent years. The first of them was
discovered by \emph{BaBar} in the reaction $e^+e^-\rightarrow
Y(4260)\: \gamma \rightarrow J/\psi \: \pi^{+} \pi^{-} \gamma$
\cite{4260}, a (narrow) state around 4260 MeV with a width of 90
MeV, later confirmed by \emph{Belle} via \emph{ISR}
\cite{Belle4260} and by \emph{CLEO} in an direct energy scan
\cite{CLEO4260}. Another state was detected at 2175 MeV by
\emph{BaBar} in the reaction $e^+e^-\rightarrow Y(2175)\: \gamma
\rightarrow \phi f_{o}(980)\gamma$ \cite{Babar2175}. \emph{Belle}
found new states at 4080, 4250, 4360, 4660 MeV in the reactions
$e^+e^-\rightarrow Y\: \gamma \rightarrow \psi(2S)\: \pi^{+}
\pi^{-}\gamma$ and $e^+e^-\rightarrow Y\: \gamma \rightarrow
J/\psi \: \pi^{+} \pi^{-}\gamma$ \cite{BelleY}. The structure of
basically all of these new states (if they will survive) is
unknown so far. 4 quark states, e. g. a $[cs][\bar{c} \bar{s}]$
state for $Y(4260)$, a $[ss][\bar{s}\bar{s}]$ state for $Y(2175)$,
hybrid and molecular structures are discussed, see also
\cite{Kalashnikova}.

\subsection{\emph{EVA-PHOKHARA} and \emph{BaBar}: \newline Baryon form factors $e^+e^-\rightarrow B \bar {B}
\gamma$}
 Also baryonic final states with protons and antiprotons,
$\Lambda^{o}$ and $\bar \Lambda^{o}$, $\Lambda^{o}$ and $\bar
{\Sigma^{o}}$, $\Sigma^{o}$ and $\bar {\Sigma^{o}}$ have been
investigated in agreement with previous measurements, but with
significantly improved statistics. The effective proton form
factor shows some nontrivial structures at invariant $p\bar{p}$
masses of 2.25 and 3.0 GeV, so far unexplained
\cite{BabarBB,Maas}.

\section{Summary}

Being discussed since the sixties/seventies $ISR$ became a
powerful tool for experimentalists only with the development of
\emph{EVA-PHOKHARA} by \emph{J. H. K\"{u}hn, H. Czy\.{z}, G.
Rodrigo} and collaborators \cite{Binner,Czyz}, a Monte Carlo
generator developed over almost 10 years, while increasing its
complexity, which is user friendly, flexible and easy to implement
into the software of the existing detectors. The $ISR$ method was
originally thought to be an alternative to energy scans to
determine hadronic cross sections from the reaction threshold up
to the maximum centre of mass energy of fixed energy meson $(\phi,
\tau/charm, B)$- factories in order to determine the hadronic
contribution to the muon magnetic anomaly $a_{\mu}^{had}$ and to
the running fine structure constant $\alpha (m_{Z}^{2})$, and it
was very successful in that respect. But being more than a
\emph{poor man's} alternative to scan energies it turned out to be
an extremely useful tool to clarify reaction mechanisms and to
reveal substructures (intermediate states and their decay
mechanisms) leading to more and more hadronic final states, a
plethora at $B$-factories. Statistical errors have been
dramatically reduced. Systematic errors are under better and
better control. Finally it has opened a totally new and unexpected
access to highly excited states (with $J^{PC} = 1^{--}$) the
structure of which is not yet known.

\end{document}